\documentclass[iop]{emulateapj}

\usepackage{lineno}
\usepackage{graphicx}
\usepackage{amsmath}
\usepackage{amssymb}
\usepackage{hyperref}

\begin{document}
\title{Millimeter Transient Point Sources in the SPTpol 100 Square Degree Survey}
\shorttitle{Transient Sources in SPTpol 100 Deg$^2$}

\def\Berkeley{1}
\def\PhysicsUChicago{2}
\def\KICPChicago{3}
\def\Dunlap{4}
\def\Cardiff{5}
\def\NIST{6}
\def\ArgonneHEP{7}
\def\FNAL{8}
\def\AAUChicago{9}
\def\EFIChicago{10}
\def\UKZN{11}
\def\SLAC{12}
\def\Caltech{13}
\def\McGill{14}
\def\CIFAR{15}
\def\ColoradoAPS{16}
\def\HarveyMudd{17}
\def\ColoradoPhys{18}
\def\UChicago{19}
\def\Stanford{20}
\def\KIPAC{21}
\def\Davis{22}
\def\LBNL{23}
\def\Michigan{24}
\def\ArgonneMSD{25}
\def\Minnesota{26}
\def\Melbourne{27}
\def\CaseWestern{28}
\def\ArtInstChicago{29}
\def\ThreeSpeedLogic{30}
\def\CfA{31}
\def\UToronto{32}
\def\illast{33}
\def\illphy{34}

\shortauthors{N.~Whitehorn, T.~Natoli, et al.}
\author{
  N.~Whitehorn\altaffilmark{\Berkeley},
  T.~Natoli\altaffilmark{\PhysicsUChicago,\KICPChicago,\Dunlap},
  P.~A.~R.~Ade\altaffilmark{\Cardiff},
  J.~E.~Austermann\altaffilmark{\NIST},
  J.~A.~Beall\altaffilmark{\NIST},
  A.~N.~Bender\altaffilmark{\ArgonneHEP,\KICPChicago},
  B.~A.~Benson\altaffilmark{\FNAL,\KICPChicago,\AAUChicago},
  L.~E.~Bleem\altaffilmark{\ArgonneHEP,\KICPChicago},
  J.~E.~Carlstrom\altaffilmark{\KICPChicago,\PhysicsUChicago,\ArgonneHEP,\AAUChicago,\EFIChicago},
  C.~L.~Chang\altaffilmark{\KICPChicago,\ArgonneHEP,\AAUChicago},
  H.~C.~Chiang\altaffilmark{\UKZN},
  H-M.~Cho\altaffilmark{\SLAC},
  R.~Citron\altaffilmark{\KICPChicago},
  T.~M.~Crawford\altaffilmark{\KICPChicago,\AAUChicago},
  A.~T.~Crites\altaffilmark{\KICPChicago,\AAUChicago,\Caltech},
  T.~de~Haan\altaffilmark{\Berkeley},
  M.~A.~Dobbs\altaffilmark{\McGill,\CIFAR},
  W.~Everett\altaffilmark{\ColoradoAPS},
  J.~Gallicchio\altaffilmark{\KICPChicago,\HarveyMudd},
  E.~M.~George\altaffilmark{\Berkeley},
  A.~Gilbert\altaffilmark{\McGill},
  N.~W.~Halverson\altaffilmark{\ColoradoAPS,\ColoradoPhys},
  N.~Harrington\altaffilmark{\Berkeley},
  J.~W.~Henning\altaffilmark{\KICPChicago,\AAUChicago},
  G.~C.~Hilton\altaffilmark{\NIST},
  G.~P.~Holder\altaffilmark{\McGill,\CIFAR},
  W.~L.~Holzapfel\altaffilmark{\Berkeley},
  S.~Hoover\altaffilmark{\KICPChicago,\PhysicsUChicago},
  Z.~Hou\altaffilmark{\KICPChicago},
  J.~D.~Hrubes\altaffilmark{\UChicago},
  N.~Huang\altaffilmark{\Berkeley},
  J.~Hubmayr\altaffilmark{\NIST},
  K.~D.~Irwin\altaffilmark{\SLAC,\Stanford},
  R.~Keisler\altaffilmark{\Stanford,\KIPAC},
  L.~Knox\altaffilmark{\Davis},
  A.~T.~Lee\altaffilmark{\Berkeley,\LBNL},
  E.~M.~Leitch\altaffilmark{\KICPChicago,\AAUChicago},
  D.~Li\altaffilmark{\NIST,\SLAC},
  J.~J.~McMahon\altaffilmark{\Michigan},
  S.~S.~Meyer\altaffilmark{\KICPChicago,\PhysicsUChicago,\AAUChicago,\EFIChicago},
  L.~Mocanu\altaffilmark{\KICPChicago,\AAUChicago},
  J.~P.~Nibarger\altaffilmark{\NIST},
  V.~Novosad\altaffilmark{\ArgonneMSD},
  S.~Padin\altaffilmark{\KICPChicago,\AAUChicago,\Caltech},
  C.~Pryke\altaffilmark{\Minnesota},
  C.~L.~Reichardt\altaffilmark{\Melbourne},
  J.~E.~Ruhl\altaffilmark{\CaseWestern},
  B.~R.~Saliwanchik\altaffilmark{\UKZN},
  J.T.~Sayre\altaffilmark{\ColoradoAPS,\ColoradoPhys},
  K.~K.~Schaffer\altaffilmark{\KICPChicago,\EFIChicago,\ArtInstChicago},
  G.~Smecher\altaffilmark{\McGill,\ThreeSpeedLogic},
  A.~A.~Stark\altaffilmark{\CfA},
  K.~T.~Story\altaffilmark{\KIPAC,\Stanford},
  C.~Tucker\altaffilmark{\Cardiff},
  K.~Vanderlinde\altaffilmark{\Dunlap,\UToronto},
  J.~D.~Vieira\altaffilmark{\illast,\illphy},
  G.~Wang\altaffilmark{\ArgonneHEP},
  and
  V.~Yefremenko\altaffilmark{\ArgonneHEP}
}

\altaffiltext{\Berkeley}{Department of Physics, University of California, Berkeley, CA, USA 94720}
\altaffiltext{\PhysicsUChicago}{Department of Physics, University of Chicago, 5640 South Ellis Avenue, Chicago, IL, USA 60637}
\altaffiltext{\KICPChicago}{Kavli Institute for Cosmological Physics, University of Chicago, 5640 South Ellis Avenue, Chicago, IL, USA 60637}
\altaffiltext{\Dunlap}{Dunlap Institute for Astronomy \& Astrophysics, University of Toronto, 50 St George St, Toronto, ON, M5S 3H4, Canada}
\altaffiltext{\Cardiff}{Cardiff University, Cardiff CF10 3XQ, United Kingdom}
\altaffiltext{\NIST}{NIST Quantum Devices Group, 325 Broadway Mailcode 817.03, Boulder, CO, USA 80305}
\altaffiltext{\ArgonneHEP}{High Energy Physics Division, Argonne National Laboratory, 9700 S. Cass Avenue, Argonne, IL, USA 60439}
\altaffiltext{\FNAL}{Fermi National Accelerator Laboratory, MS209, P.O. Box 500, Batavia, IL 60510}
\altaffiltext{\AAUChicago}{Department of Astronomy and Astrophysics, University of Chicago, 5640 South Ellis Avenue, Chicago, IL, USA 60637}
\altaffiltext{\EFIChicago}{Enrico Fermi Institute, University of Chicago, 5640 South Ellis Avenue, Chicago, IL, USA 60637}
\altaffiltext{\UKZN}{School of Mathematics, Statistics \& Computer Science, University of KwaZulu-Natal, Durban, South Africa}
\altaffiltext{\SLAC}{SLAC National Accelerator Laboratory, 2575 Sand Hill Road, Menlo Park, CA 94025}
\altaffiltext{\Caltech}{California Institute of Technology, MS 249-17, 1216 E. California Blvd., Pasadena, CA, USA 91125}
\altaffiltext{\McGill}{Department of Physics, McGill University, 3600 Rue University, Montreal, Quebec H3A 2T8, Canada}
\altaffiltext{\CIFAR}{Canadian Institute for Advanced Research, CIFAR Program in Cosmology and Gravity, Toronto, ON, M5G 1Z8, Canada}
\altaffiltext{\ColoradoAPS}{Department of Astrophysical and Planetary Sciences, University of Colorado, Boulder, CO, USA 80309}
\altaffiltext{\HarveyMudd}{Harvey Mudd College, 301 Platt Blvd., Claremont, CA 91711}
\altaffiltext{\ColoradoPhys}{Department of Physics, University of Colorado, Boulder, CO, USA 80309}
\altaffiltext{\UChicago}{University of Chicago, 5640 South Ellis Avenue, Chicago, IL, USA 60637}
\altaffiltext{\Stanford}{Dept. of Physics, Stanford University, 382 Via Pueblo Mall, Stanford, CA 94305}
\altaffiltext{\KIPAC}{Kavli Institute for Particle Astrophysics and Cosmology, Stanford University, 452 Lomita Mall, Stanford, CA 94305}
\altaffiltext{\Davis}{Department of Physics, University of California, One Shields Avenue, Davis, CA, USA 95616}
\altaffiltext{\LBNL}{Physics Division, Lawrence Berkeley National Laboratory, Berkeley, CA, USA 94720}
\altaffiltext{\Michigan}{Department of Physics, University of Michigan, 450 Church Street, Ann  Arbor, MI, USA 48109}
\altaffiltext{\ArgonneMSD}{Materials Sciences Division, Argonne National Laboratory, 9700 S. Cass Avenue, Argonne, IL, USA 60439}
\altaffiltext{\Minnesota}{School of Physics and Astronomy, University of Minnesota, 116 Church Street S.E. Minneapolis, MN, USA 55455}
\altaffiltext{\Melbourne}{School of Physics, University of Melbourne, Parkville, VIC 3010, Australia}
\altaffiltext{\CaseWestern}{Physics Department, Center for Education and Research in Cosmology and Astrophysics, Case Western Reserve University, Cleveland, OH, USA 44106}
\altaffiltext{\ArtInstChicago}{Liberal Arts Department, School of the Art Institute of Chicago, 112 S Michigan Ave, Chicago, IL, USA 60603}
\altaffiltext{\ThreeSpeedLogic}{Three-Speed Logic, Inc., Vancouver, B.C., V6A 2J8, Canada}
\altaffiltext{\CfA}{Harvard-Smithsonian Center for Astrophysics, 60 Garden Street, Cambridge, MA, USA 02138}
\altaffiltext{\UToronto}{Department of Astronomy \& Astrophysics, University of Toronto, 50 St George St, Toronto, ON, M5S 3H4, Canada}
\altaffiltext{\illast}{Astronomy Department, University of Illinois at Urbana-Champaign, 1002 W. Green Street, Urbana, IL 61801, USA}
\altaffiltext{\illphy}{Department of Physics, University of Illinois Urbana-Champaign, 1110 W. Green Street, Urbana, IL 61801, USA}
\email{nwhitehorn@berkeley.edu}
\email{t.natoli@utoronto.ca}


\begin{abstract}
The millimeter transient sky is largely unexplored, with measurements limited to follow-up of objects detected at other wavelengths.
High-angular-resolution telescopes designed for measurement of the cosmic microwave background offer the possibility to discover new, unknown transient sources in this band, particularly the afterglows of unobserved gamma-ray bursts.
Here we use the 10-meter millimeter-wave South Pole Telescope, designed for the primary purpose of observing the cosmic microwave background at arcminute and larger angular scales, to conduct a search for such objects.
During the 2012--2013 season, the telescope was used to continuously observe a 100 $\deg^2$ patch of sky centered at RA 23\textsuperscript{h}30\textsuperscript{m} and declination -55$^\circ$ using the polarization-sensitive SPTpol camera in two bands centered at 95 and 150 GHz.
These 6000 hours of observations provided continuous monitoring for day- to month-scale millimeter-wave transient sources at the 10~mJy level.
One candidate object was observed with properties broadly consistent with a gamma-ray burst afterglow, but at a statistical significance too low ($p=0.01$) to confirm detection.
\end{abstract}

\maketitle

\section{Introduction}

Millimeter-wave observations of variable and transient astrophysical sources have contributed greatly to our understanding of the processes in these objects, for example through observations of outbursts from active galactic nuclei (e.g. \citealt{dent83}) and the detection of reverse shocks in gamma-ray bursts \citep{laskar13}.
Gamma-ray burst (GRB) afterglows are of particular interest in this band as they often have the peak of their spectra in or near the millimeter range \citep{granot02}, with emission lasting over timescales of days to weeks.
As GRB emission is expected to be more tightly beamed in gamma rays than at longer wavelengths, burst afterglows not accompanied by detectable gamma ray emission are believed to exist but have not been detected.
The observation of these off-axis sources would provide insight into the jet dynamics and central engine energy budget of GRBs \citep{rhoads97}.
In addition, other classes of gamma-dark bursts have been advanced as the solution to a number of astrophysical puzzles, for example the origin of the TeV--PeV diffuse neutrino background \citep{senno16}.
However, no untriggered millimeter transient searches---which could reveal both these orphan GRB afterglows and new, unknown sources---have been conducted to date due to limitations of observing time and field of view.

High-angular-resolution cosmic microwave background (CMB) surveys offer a unique opportunity to fill this void and to probe for previously unknown transient sources in the millimeter and submillimeter bands \citep{metzger15}.
To average down instrumental and atmospheric noise, a typical ground-based CMB survey will continuously scan the same patch of sky (tens to thousands of square degrees) for years.
In addition to providing low-noise maps of the cosmic microwave background, this observation strategy provides a platform for continuous monitoring of the survey region for variable and transient sources in the millimeter band in which the CMB is brightest.
The rapid reobservation cadence of ground-based instruments, typically hours, provides sensitivity to a wide range of possible variability scales, from hours to the years-long periods of the cosmology surveys.

This work describes such a search for transient point sources using the 10-meter South Pole Telescope \citep[SPT; ][]{carlstrom11, austermann12}.
Using the SPT, we achieve discovery sensitivity of approximately 10~mJy on timescales of days to weeks (Sec.~\ref{sec:sensitivity}).
This depth gives a sensitivity in the upper range of observed GRB afterglows from follow-up observations conducted in this band, but well below the brightest observed bursts, which have had fluxes exceeding 70~mJy \citep{postigo12}.
This sensitivity also compares favorably to previous blind radio surveys (e.g.~\citealt{levinson02,bell11}), which have had comparable flux sensitivity to this work but smaller effective sky coverage for week-scale sources and have been conducted at lower frequencies where GRB afterglows are much dimmer.

\section{Survey Method}
\label{sec:method}

\begin{figure}
\includegraphics[width=\linewidth]{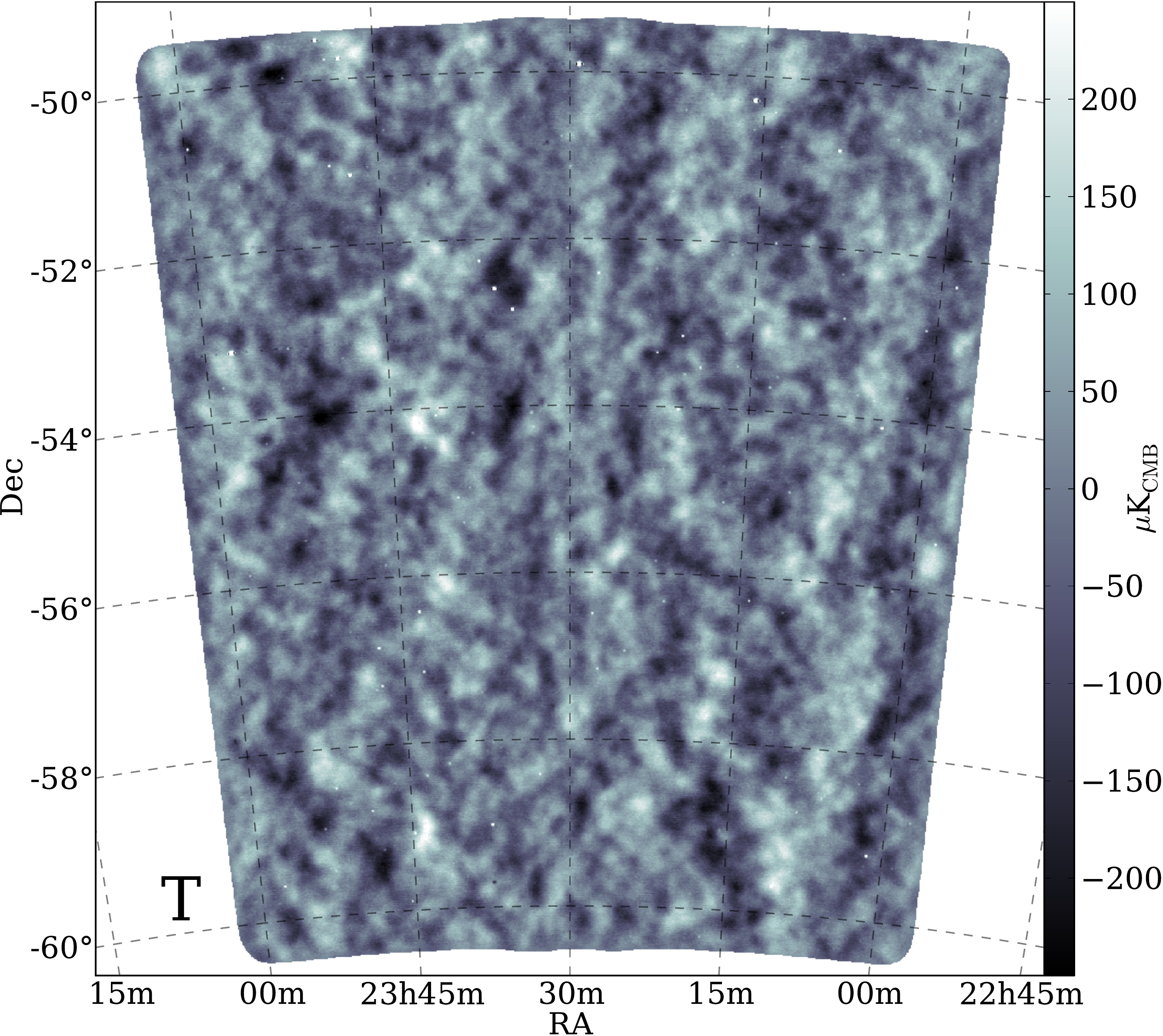}
\caption{Survey field used for the SPTpol observations in this work, showing surface brightness (T) at 150~GHz. This field is shared with recent SPTpol cosmology analyses (e.g. \citealt{keisler15}) and is well out of the Galactic plane.}
\label{fig:field}
\end{figure}

Over the period of this study (2012 April -- 2013 May), the polarization-sensitive SPTpol receiver was used to observe a 100 square degree field centered at RA 23\textsuperscript{h}30\textsuperscript{m} and declination -55$^\circ$ (Fig.~\ref{fig:field}).
This field is well out of the Galactic plane, giving sensitivity predominantly to extragalactic sources.
Observations were conducted continuously at approximately hourly intervals except for the period from 2012 November 10 to 2013 March 26, in which observations of this field were not made to avoid sun contamination and to allow for telescope maintenance.

The SPTpol receiver consists of an array of 1536 transition edge sensor bolometers, with 360 operating at 95~GHz and 1176 at 150~GHz.
Each of the receiver's pixels consists of a pair of bolometers sensitive to perpendicular linear polarizations at one of these frequencies \citep{austermann12}.
The 95~GHz detectors have wider beams (1.8 arcmin) than the 150~GHz detectors (1.1 arcmin).
For unresolved sources with comparable fluxes at both frequencies, such as GRB afterglows, the change in beam width as well as differing detector performance, the larger number of 150~GHz detectors, and changes in the atmospheric noise level result in 1.7 times better sensitivity at 150~GHz.
Much steeper spectra than expected for GRBs---steeper than $\nu^{-1.2}$---would be required for better sensitivity at 95~GHz.
In this work, we focus on sources with less steeply falling or rising spectra and use the 150~GHz band as our primary detection channel, examining the 95~GHz data only for additional information about any detected candidate sources.

The cryogenic system used to maintain the SPTpol detectors at sub-Kelvin temperatures is a closed cycle three-stage ($^4$He-$^3$He-$^3$He) refrigerator.
The refrigerator's cooling cycle is not continuous and needs to be recycled periodically for recondensation of helium.
The recycling causes approximately 8 hours of lost observation time every 36 hours as the helium is recondensed, the detectors retuned, and the instrument recalibrated.
As the transient search conducted here is focused on emission lasting days or longer, we combine all field observations within one of these 36-hour cycles into a single map for analysis using an inverse-variance-weighted average. 
Because the instrument calibration and detector operating points are maintained for this period, this also results in the combination of maps with similar calibration and noise properties.
The resulting 150~GHz cycle-length maps have a median $1 \sigma$ depth of 5.0~mJy.

These cycle-length combined maps we then filter, as described in the following sections, and compare to the average map over the full season. 
Using a multi-epoch likelihood method, we examine these maps for sources present for periods of time significantly shorter than the one-year survey period.


\subsection{Data Filtering and Calibration}

Atmospheric fluctuations produce large-angular-scale signals in the data that vary from observation to observation and must be filtered to remove false transient sources.
Here we combine two approaches for the removal of these signals: filtering time-ordered detector data against large-scale variations and a matched filter applied to the maps to increase sensitivity to point sources.
During observations, the telescope scans back and forth across the observing field in azimuth (which is equivalent to right ascension at the South Pole) at a speed of $\sim 0.5^\circ\, \mathrm{s}^{-1}$ and then steps in elevation (equivalent to declination). 
The telescope repeats this scanning and stepping over the declination range of the survey field.
As a first step in the filtering of large-scale map structures produced by atmospheric fluctuations, we subtract a seventh-order polynomial fit to the data from each $\sim 10^\circ$ azimuthal scan.
This filtering removes most atmospheric structure at degree and larger scales while having little effect on sensitivity to point sources at arcminute scales.
The time-ordered detector data are then low-pass filtered at a frequency corresponding to the 0.25 arcmin pixels in the final maps to prevent aliasing higher frequency features.
After filtering, the data from each detector is combined into a map using an equal-area projection.
Both polarizations are summed and the contributions from each detector weighted by the inverse of its variance in the 0.8-3 Hz band, roughly where we expect the signals of interest to most SPTpol analyses to lie.

As time-independent features (the CMB, steady point sources, etc.) difference out in our analysis (Sec.~\ref{subsec:flaresearch}), our map filtering is built to discriminate point sources from the time-varying features of the sky: atmospheric fluctuations, instrument noise, and variable compact objects.
We treat the first two of these using the matched filter method described in \cite{vieira10} and \cite{haehnelt96} to downweight remaining large-scale structures in the map in favor of the instrument beam scale features produced by point sources.
The matched filter is constructed primarily from measurements of the instrument beam and data-driven estimates of the instrument and atmospheric noise power spectra.
Although not required for this search, we also include the time-independent CMB power spectrum as a noise term in the matched filter for consistency with the previous SPT point source results in \cite{vieira10} and \cite{mocanu13}.
As the typical angular scale of CMB fluctuations is much larger than the instrument beam, filtering the CMB removes little of the point source signal and thus does not significantly degrade our analysis.

Many of the brightest point sources at millimeter wavelengths exhibit substantial time variability---up to a factor of two for some active galactic nuclei (AGN).
Our observing field contains $\sim 350$ point sources detected with fluxes above 2.5~mJy ($5 \sigma$ for the year of data used here).
To prevent false detections from variability in these sources, we mask areas of the map within 5 arcmin of  known steady point sources in this field from earlier SPT results \citep{mocanu13} with quiescent fluxes above 5~mJy, well below our threshold for detection of shorter-duration transient sources (Sec.~\ref{sec:sensitivity}).
Very bright sources produce detectable filtering wings at larger distances that alias fluctuations in the source intensity.
As such, we extend the masked area of sources above 50~mJy to a 10 arcmin radius from the source.
This leaves a final survey array of $80.5\,\textrm{deg}^2$.
These sources were also masked in the computation of the polynomials subtracted from the time-ordered detector data.

Overall calibration of our data is based on observations of the galactic H~\textsc{II} region RCW38 for flux calibration and coarse pointing, along with planet observations and AGN for measurements of instrument beams and pointing.

\subsection{Data Selection}
\label{subsec:selection}

We use nearly identical observation quality criteria as in \cite{keisler15}.
These criteria remove observations of the field with elevated or non-Gaussian noise as well as periods with abnormally low observing efficiency due, for example, to telescope maintenance or hardware problems.
In addition, we use a different time discretization that causes a further 1\% loss of observing time.
This leaves 253 days of sensitivity to week-scale emitting sources.

\subsection{Flare Identification}
\label{subsec:flaresearch}

To identify transient sources, we used a multi-epoch method following \cite{braun10} sensitive to sources on all time scales from the map discretization ($\sim 36$ hours) to a few months.
This method identifies sources by fitting a variable-width flare template to the inferred point source flux at a particular position as a function of time.
We then use the likelihood ratio of this fit to the null hypothesis (zero peak flux) as a discriminant to identify potential sources.
For each point on a grid covering the survey area with resolution of half an arcminute (approximately half our beam size), we minimize the following over the time series at that point:

\begin{equation}
-2 \ln \mathcal{L} (S, t_0, w) = \sum_t \frac{\left (\phi_t - f(t; S, t_0, w) \right)^2}{\sigma_t^2} - 2 P(w).
\label{eq:llh}
\end{equation}

Here $\phi_t$ is the difference between each pixel and that pixel in the full-period average map, $\sigma_t$ is the estimated noise level at that map position and time, $P(w)$ is a penalty function that will be described later, and $f(t; S, t_0, w)$ is a Gaussian template for the source flux as a function of time:

\begin{equation}
f(t; S, t_0, w) = S e^{-(t - t_0)^2 / (0.25 w^2 / \ln 2)}.
\label{eq:gaussprofile}
\end{equation}

The functional form of this template was chosen as a generic search function containing a variable-width flare that allows the numerically robust minimization of equation~\eqref{eq:llh} and provides good statistical power for a wide variety of potential flare shapes \citep{braun10}.
The parameter $w$ is the full width at half-maximum (FWHM) of this Gaussian.
The sensitivity to astrophysical sources identified using this method depends weakly, typically at the percent level, on the actual emission profile.
This can be seen intuitively by considering the limiting cases of low and high signal-to-noise.
In the low signal-to-noise regime, at the detection threshold, the data would be sufficient to detect a source but not to determine the shape of the emission.
This is equivalent to the statement that there cannot be a large change in the likelihood \eqref{eq:llh}, our detection figure of merit, from variations in the functional form of the emission profile $f(t; \boldsymbol x)$.
When signal-to-noise is very high, we do expect potentially large changes in the value of the likelihood from shape mismatches.
However, as our detection threshold (section~\ref{sec:sensitivity}) is fairly low, we do not expect changes in the likelihood from shape differences relative to our template to meaningfully change our detection efficiency.
The major impact of shape mismatches instead is to cause the parameter values $(S, t_0, w)$ to reflect only effective parameters of our template rather than unbiased estimates of the peak flux, start time, or emission width.
As genuine astrophysical sources are not expected to have a Gaussian profile, the parameters of equation~\eqref{eq:gaussprofile} should thus be regarded in general as nuisance parameters.

From equation~\eqref{eq:llh}, we form a test statistic (TS) from the ratio of the best-fit likelihood with all parameters free to the best-fit likelihood where the estimated peak flux is fixed to $S = 0$ and all other parameters are free (here carets denote best-fit quantities):

\begin{equation}
\textrm{TS} = -2 \Delta \ln \mathcal L = -2 \ln \mathcal{\hat L}(\hat S, \hat t_0, \hat w) + 2 \ln \mathcal{\hat L}(0, \hat t_0 ', \hat w ').
\label{eq:ts}
\end{equation}

If the penalty function $P(w) = 0$, this test statistic has a maximization bias to short estimated flare widths ($w$) as a result of the look-elsewhere effect.
For short flares, there are more potential uncorrelated points in time for the flare to start ($t_0$), effectively widening the search space and thus the effective trials factor.
This increases the false discovery rate (FDR) at small $w$.
Following \cite{braun10}, we flatten the FDR by applying a penalty term $P(w) = \ln(w)$ that approximates a marginalization of the likelihood \eqref{eq:llh} over a uniform prior in $t_0$ and cancels this effect.

For the evaluation of $P(w)$, $w$ is bounded above by the full length of the survey to prevent runaway values of equation~\eqref{eq:llh} when evaluating the null hypothesis ($S = 0$), in which $P(w) = \ln(w)$ is the only variable term.
$w$ is otherwise constrained only by a non-negativity requirement.

\begin{figure}
\includegraphics[width=\linewidth]{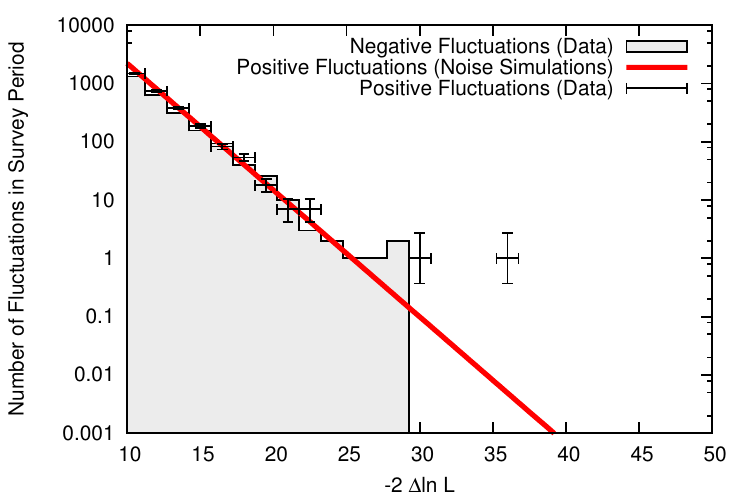}
\caption{Distribution of test statistic values obtained in this analysis. The red line is a smoothed version of $2 \times 10^9$ Monte Carlo realizations of the estimated noise in the field and describes the data (crosses) well in the noise-dominated region at the left. The gray filled region shows the values for the negative fluctuations observed in our data, which are unphysical as source emission and which we use as a control sample. Vertical error bars are 68\% Feldman-Cousins confidence intervals; horizontal error bars indicate bin width. The point at far right is described in Sec.~\ref{sec:source}.}
\label{fig:tshist}
\end{figure}

We compute statistical significance from values of TS by using the noise-dominated low-significance parts of the TS distribution and by using negative fluctuations, which are unphysical as source emission, as a signal-free control sample (Fig.~\ref{fig:tshist}).
Both distributions, with the exception of the object described in section \ref{sec:source}, are well described in their high-significance region by the expected $\chi^2$ distribution.
Above the highest observed noise fluctuation, we extrapolate this fit distribution of positive fluctuations to more signal-like values of TS to compute significance.
We use the significance of the highest-TS point as a summary statistic for the entire analysis, resulting in our final $p$-value being equal to the $p$-value for the highest significance point.
In the limit that the false detection rate is small ($\ll 1$), this is turn is equal to the false detection rate associated with the most significant point in the survey.

As a further cross-check, we ran $2 \times 10^9$ noise-only simulations, equivalent to approximately 1500 years of observations on this field.
These reproduce the data well in the noise-dominated part of the TS distribution (Fig.~\ref{fig:tshist}) and show no evidence of deviations from the $\chi^2$ extrapolation at high values.

\section{Sensitivity}
\label{sec:sensitivity}

\begin{figure}
\includegraphics[width=\linewidth]{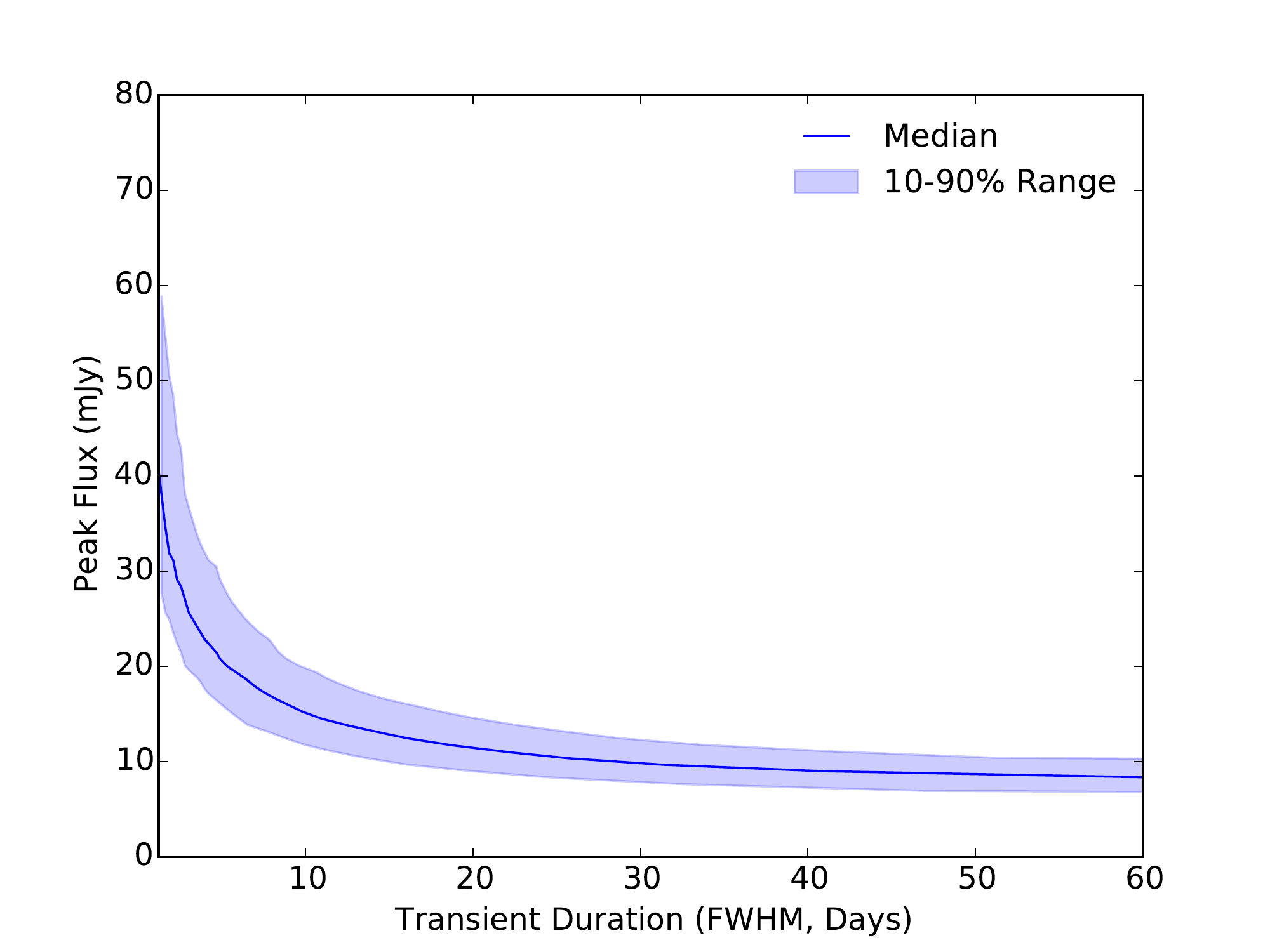}
\caption{Detection threshold ($6 \sigma$) as a function of flare duration.
The blue line indicates the flux at which 50\% of sources would be detectable; the filled region shows the range between the points at which 10\% of sources and at which 90\% of sources would be detectable.
Very long flares, approaching the length of the full dataset, are indistinguishable from continuously-emitting point sources, reducing sensitivity slightly relative to $t^{-1/2}$ on the right.
This figure assumes an arbitrary Gaussian flare profile---alternatives (boxcar functions, scaled copies of the GRB030329 100~GHz lightcurve from \citealt{sheth03}) differ at the few percent level.
Sources were injected starting at times uniformly distributed through the observing period within 2 days of an observation (this removes the summer maintenance period, leaving 253 days of effective live time) and uniformly throughout the 80.5 deg$^2$ masked survey area.
}
\label{fig:gausssens}
\end{figure}

Given the one-year length of the survey, observing band, and flux sensitivity (Fig.~\ref{fig:gausssens}), the objects most likely to be detected are expected to be nearly on-axis GRBs, tidal disruption events, and blazar flares \citep{metzger15}.
The source class with the highest predicted rate from unknown sources (i.e. neglecting flares from AGN whose quiescent flux is above SPT's threshold) is the nearly on-axis GRB a few days to a week after the burst \citep{ghirlanda13,metzger15}.

GRB afterglow emission is believed to be dominated by synchrotron processes, with significant self-absorption at low frequencies.
As the afterglow ages, it gradually becomes optically thin and the self-absorption point ($\nu_a$) moves to progressively lower frequencies.
This break frequency marks the junction between the rising optically thick part of the spectrum ($\nu^{2-2.5}$) and the falling optically thin synchrotron regime ($\nu^{-\beta}$) and thus corresponds to the peak of the spectrum.
In the early stages of the afterglow, the burst will brighten with time at observing frequencies $\nu_\mathrm{obs}$ below $\nu_a$ as $\nu_a$ moves to lower frequencies faster than the burst cools, decreasing the suppression from self-absorption at $\nu_\mathrm{obs}$.
Once $\nu_a < \nu_\mathrm{obs}$, the flux will begin to decrease with time, following the cooling of the burst.
This competition between cooling and self-absorption results in an earlier peak emission time (when $\nu_a = \nu_\mathrm{obs}$) as $\nu_\mathrm{obs}$ rises, corresponding to both brighter peak emission and tighter beaming angles.
For $\nu_\mathrm{obs} \sim 150$~GHz, typical peak times are a few days to a week after the burst rather than the several weeks typical of 1.4~GHz observations, with much higher peak fluxes \citep{ghirlanda13}.
This peak emission time, and the observed length of emission, may be correlated with the peak observed flux both due to the physics of the expanding jet and due to cosmological effects such as time dilation at high redshift \citep{ghirlanda13}.
Although the tighter beaming angles at high frequencies suppress the number of observable bursts, the higher fluxes result in a net increase in the number of detectable objects for a survey with a fixed limiting flux density as $\nu_\mathrm{obs}$ increases into the millimeter band \citep{ghirlanda14,metzger15}.

The approach described in Sec.~\ref{sec:method} gives effective $1 \sigma$ map noise of 2--3~mJy on the relevant week timescale for on-axis GRB afterglows.
For the longer (month-scale) emission expected from very off-axis and population-three bursts \citep{macpherson15,ghirlanda14}, we achieve lower effective noise in the 1--2~mJy range
At very long timescales ($\gtrsim$ 6 months), however, sensitivity rapidly fades as the source duration becomes comparable to the survey period and it becomes indistinguishable from a steady source.

Over the number of map pixels and time range of the survey, we expect to have up to $5\sigma$ (TS=25) fluctuations by chance (red line, Fig.~\ref{fig:tshist}).
This makes $6\sigma$ (TS=36), which corresponds to 0.01 false detections in the survey, a reasonable benchmark for detection for the purpose of computing sensitivity.
At the $6\sigma$ level, this gives an average detection sensitivity of peak fluxes in the 10--15~mJy range (Fig.~\ref{fig:gausssens}), depending on emission length and position in the field.

This sensitivity is well below the brightest GRBs followed up in this band, which had peak fluxes above 70~mJy (e.g. GRB030329 from \citealt{sheth03}), but well above the average observed burst, which has a peak flux of $\sim 1$~mJy.
Using the catalog in \cite{postigo12}, we would have been sensitive to $\sim 6\%$ of the bursts with measurements---either limits or detections---comparable to our sensitivity.
This is not an unbiased catalog, however, so the implications of this for the true average burst are not entirely clear.
Whether we assume this to be a representative sample or use theoretical calculations such as \cite{metzger15} or \cite{ghirlanda14}, we expect only a small number ($\lesssim 1$) of detectable bursts in the survey area per year.
This expected number depends on the GRB jet opening angle and a number of other poorly known parameters and so is not well-determined theoretically.

\section{Candidate Object}
\label{sec:source}

\begin{figure}
\includegraphics[width=\linewidth]{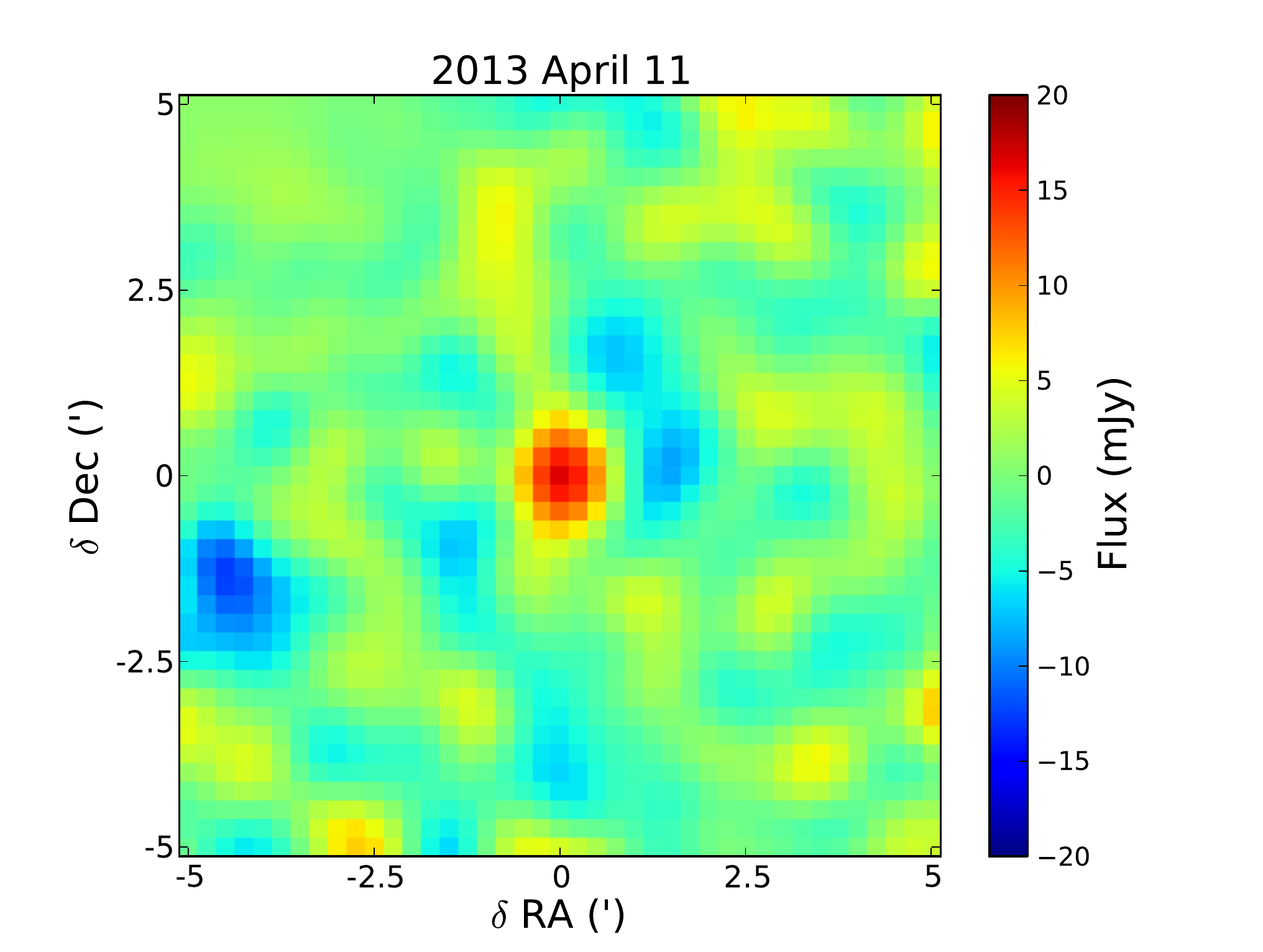}
\includegraphics[width=\linewidth]{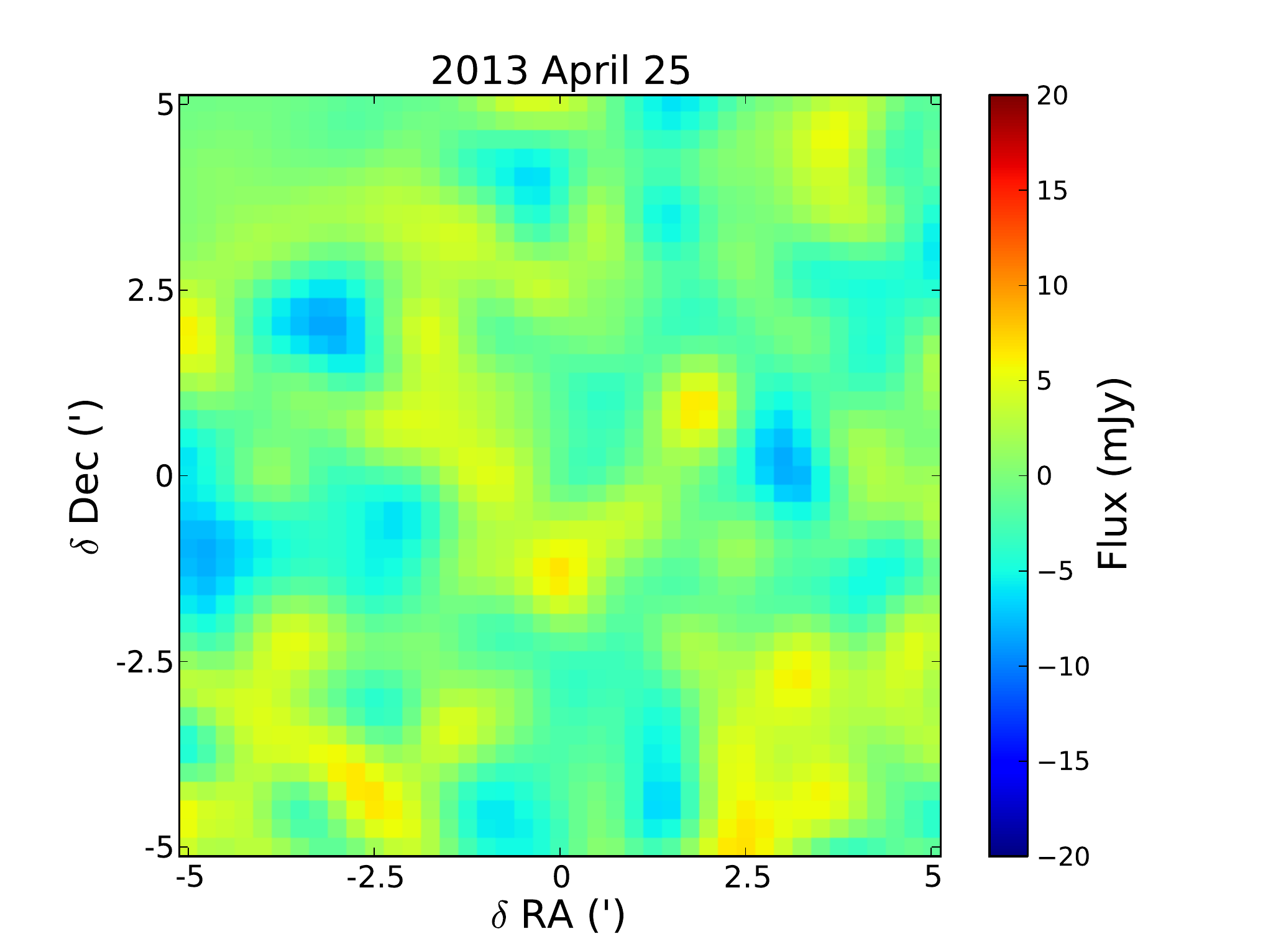}
\caption{Filtered maps of the region around the candidate source at 150~GHz at the peak time (2013 April 11) and off-time two weeks later (2013 April 25).
The color scales are identical in both panels.
Pixels are 0.25 arcminutes across.
The telescope has a 1.1 arcminute beam at this frequency.
The $1 \sigma$ noise in this field is 3.3~mJy for both figures.
These maps correspond to fridge-cycle-length time slices of approximately 36 hours; the top panel shows the same time period as the peak point in Fig.~\ref{fig:sourcelc}, while the bottom panel corresponds to the second point from the right.
}
\label{fig:sourcemap}
\end{figure}

\begin{figure}
\includegraphics[width=\linewidth]{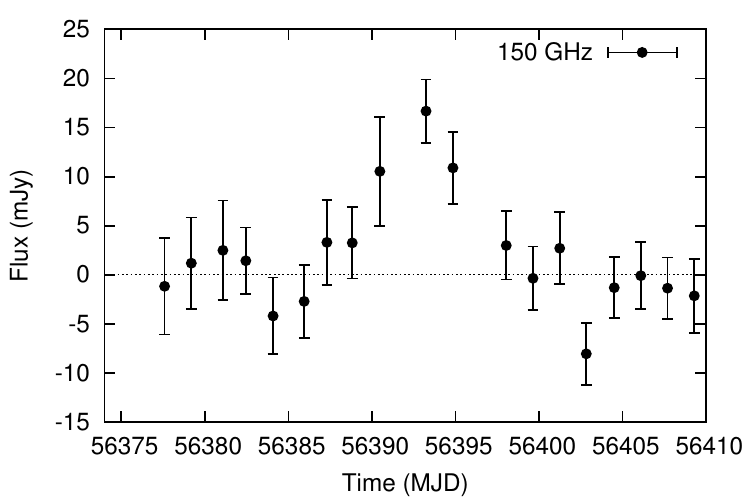}
\caption{Flux vs. time around the peak time of the candidate source at 150~GHz.  Points are placed at the start of the data taking period to which they correspond. Error bars reflect $1\sigma$ uncertainties.}
\label{fig:sourcelc}
\end{figure}

One candidate object was observed peaking on 2013 April 11 at 23\textsuperscript{h}52\textsuperscript{m}30\textsuperscript{s}, $-57^\circ30'7''$ (J2000), with a best-fit peak flux at 150~GHz of $16.5 \pm 2.4$~mJy (Fig.~\ref{fig:sourcemap}) and emission above background levels detected for three days on either side of the peak ($\hat w = 6.3$~days, Fig.~\ref{fig:sourcelc}).
Using the statistical significance calculation from Sec.~\ref{subsec:flaresearch}, 0.007 objects of this TS value (37) or higher were expected by chance in this dataset.

The data contributing most to the likelihood show no signs of data quality problems.
During the peak times of the candidate source on April 11, there is excess flux at this location in 16 of the 20 individual one-hour maps bundled for the analysis.
In nine of these one-hour maps, the excess is more than $1\sigma$, in four, more than $2\sigma$, and in one, $3\sigma$.
No single map contributes more than this, which implies that no single observation dominates the observed excess.
This rules out the kinds of brief instrumental systematics identified in previous radio transient surveys in \cite{frail2012}.

Another potential systematics issue arises from a day-long power outage on April 9--10 that stopped telescope observations during this period. 
Detailed data quality checks on maps from the cycle beginning April 11, after the outage, showed no evidence of data quality problems (non-Gaussian noise, higher than normal noise levels, shifts in position or flux of steady in-field sources) induced by the outage.

In the 95~GHz band, no corresponding source was observed at this time.
Using the best-fit values of $w$ and $t_0$ from 150~GHz, we can set a limit on the 95 GHz peak flux of $S_{95} < 8$~mJy at 90\% CL, corresponding to a constraint on the spectral index of $\alpha > 1.5$ ($F_\nu \propto \nu^\alpha$) at 90\% CL.
This is consistent either with the candidate object being a statistical fluctuation or with a strongly inverted spectrum.
As thermal emission at this level over a short period of time is unlikely, such a spectrum would be best explained by self-absorbed synchrotron emission with a cutoff above 100~GHz ($\alpha = 2-2.5$).
GRB afterglows in this frequency range are expected to have self-absorption cutoffs in the millimeter and submillimeter band and rising ($\alpha \sim 2$) or weakly inverted ($\alpha = 0.3$) spectra at 100~GHz \citep{granot02}.

\begin{figure}
\includegraphics[width=\linewidth]{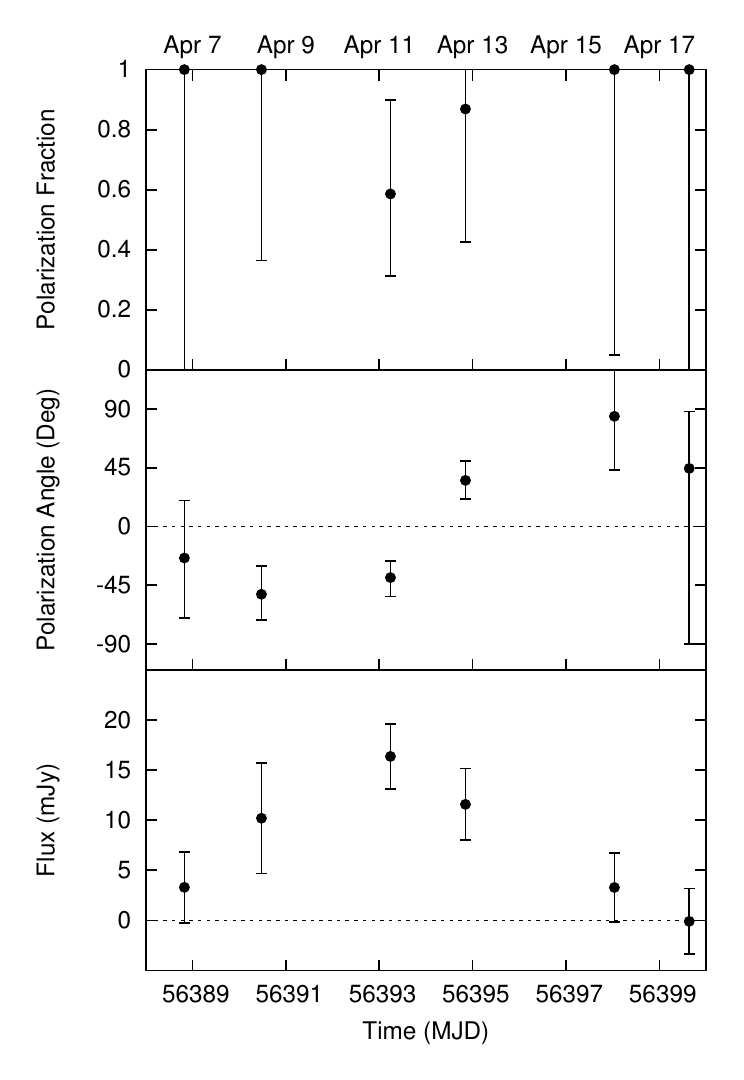}
\caption{Polarization properties of the candidate source in the 150~GHz band as a function of time. Quoted uncertainties are approximate $1\sigma$ errors and represent points in profile likelihood space at which $\Delta 2 \ln \mathcal{L} = 1$. The average polarization fraction in this period is $77 \pm 23 \%$. Points are placed at the start of the data taking period to which they correspond. A subset of Fig.~\ref{fig:sourcelc} is included for reference.}
\label{fig:polarization}
\end{figure}

The 150~GHz emission detected at the peak time was highly linearly polarized, with a polarization fraction $f_p = 0.6 \pm 0.3$ (SPTpol is not sensitive to circular polarization).
Similar polarization fractions were observed at every point in the light curve with signal-to-noise in flux greater than one (Fig.~\ref{fig:polarization}).
This is consistent with emission from a small-volume synchrotron source such as a GRB or other small jet produced in an extremely homogeneous magnetic field \citep{granot03}.
It is not clear, however, how this very high polarization fraction corresponds to the non-detection in the 95~GHz data, which suggests an optically thick source.

Up to and including the peak of the emission on April 11, the detected polarization angle was consistent within statistical uncertainties, although only two points (April 9 and 11) have uncertainties small enough ($10^\circ$) to draw any conclusions on this point.
Beginning on April 12, as the candidate source began to fade, the polarization angle abruptly rotated $75 \pm 15^\circ$, while maintaining a high polarization fraction of $0.8^{+0.2}_{-0.4}$.
No further changes to the polarization fraction or angle were observed after that point, although the subsequent rapid reduction in flux makes any determination of polarization properties after April 15 difficult.
Averaging all maps before the shift on April 12, the detected mean polarization angle of emission up to the peak was $-42 \pm 10^\circ$.
After the shift, the mean angle rotated to $36 \pm 14^\circ$.
Such a $\sim 90^\circ$ polarization angle rotation at this point in the lightcurve would be typical of a GRB afterglow jet break \citep{granot03,wiersema2014}, in which the beaming angle expands beyond the geometric opening angle of the jet and the viewable polarization field becomes truncated.

We determine the overall mean polarization fraction of the emission using a profile likelihood, in which we form contours in an assumed constant polarization fraction $\langle f_p \rangle$, with all other parameters (polarization angle, true source flux) optimized to their best-fit values for each value of $\langle f_p \rangle$ and allowed to vary without constraint in time.
The difference in this profile likelihood between the best-fit point ($\langle f_p \rangle = 0.77 \pm 0.23$) and $f_p = 0$ ($2 \Delta \ln \mathcal L = 13$) lets us test for the statistical significance of the detection of non-zero polarization.
As atmospheric foregrounds are unpolarized and instrumental polarization leakage is low \citep{keisler15}, the detection of a polarization fraction $f_p > 0$ can be taken as independent \textit{a posteriori} evidence for an astrophysical source.
Using Monte Carlo simulations, we would have expected to have a polarization $2 \Delta \mathcal L \ge 13$ by chance in these data given the observed intensity curve in $1.5\%$ of cases.
As $f_p$ is statistically independent of $S$, and no selection was performed on $f_p$, this significance does not require correction for the look-elsewhere effect.
As a systematics check, we examined the apparent polarization fraction at the locations of lower-significance fluctuations in the maps containing the candidate source.
We found no evidence for correlation between intensity and polarization fraction at these points, ruling out a temporary miscalibration or other systematic source of polarization in the data.

No steady sources at this location have been observed in SPT data and there was no evidence for emission at other times in this survey (Fig.~\ref{fig:sourcemap}, 150~GHz quiescent flux $< 1.3$~mJy at 90\% CL).
Although several known sources are present within the half-arcminute positional uncertainty on this candidate, data from the Blanco Cosmology Survey and Spitzer (taken before the potential flare date) show no bright or otherwise notable sources at this location that would indicate a likely counterpart \citep{ashby13, bleem15}.
One dim catalogued GALEX source \citep{galexGR6} is present at this position, though again the source density is high enough to prevent a definite association.
The absence of a bright source in these surveys suggests that the observed emission was likely not a minor flare-up of an AGN with quiescent flux just below our threshold.

Due to its high galactic latitude ($b=-58^\circ$), the candidate is unlikely to be a galactic source.
If the candidate is an extragalactic source, the absence of a bright catalogued host galaxy at this position would imply a high ratio of source to host luminosity.
No gamma-ray or X-ray alerts were filed to GCN from this region of the sky within several weeks of this event, and no alerts at any time in 2012 or 2013 were consistent with this position.

Although the statistical significance of this event in our analysis is low ($p = 0.01$), the polarization data provide an independent, though not conclusive, chain of evidence in support of the idea that these observations were due to some astrophysical transient.
The nature of that potential transient remains unclear, as no additional information from the 95~GHz band or other observations provide a positive spectral measurement, counterpart, or host galaxy.
Although it is not clear how to reconcile the 95~GHz non-detection with the 150~GHz polarization data, a plausible explanation for the 150~GHz data alone would be a nearly on-axis GRB afterglow.
This would be consistent with the timescale of emission, the high degree of linear polarization, and the $90^\circ$ polarization rotation coincident with the beginning of the candidate's decay \citep{granot03}.
The gamma-ray component of such a GRB could have been missed due to the limited observing efficiency of satellites, a small misalignment of the jet, or obscuration of the prompt high-energy component.
The last two would be consistent with the several-day rise in emission seen here.

\section{Population Constraints}

Beyond the properties of the highest-significance point, the distribution of TS values (Fig.~\ref{fig:tshist}) allows us to place constraints on the population of sources.
A sufficiently steep distribution dominated by dim sources would have introduced statistical non-Gaussianity in the maps by increasing the rate of subthreshold positive fluctuations above expected levels, whereas a relatively flat spectrum of peak fluxes would be expected to produce a uniform distribution in $\sqrt{\mathrm{TS}}$.

\begin{figure}
\includegraphics[width=\linewidth]{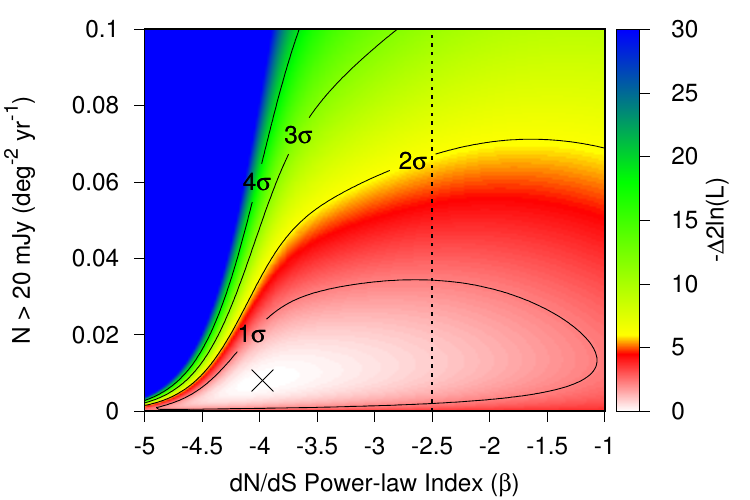}
\caption{Constraints on the distribution of source brightnesses for one-week (FWHM) Gaussian sources.
Sources were injected using the same procedure as in Fig.~\ref{fig:gausssens} with $dN/dS \propto S^\beta$ and the resulting test-statistic distribution compared to the data.
The $\times$ indicates the best-fit point in the parameter space, corresponding to one expected detection in the SPTpol 100 square degree survey.
Labeled exclusion significances are computed from the likelihood ratio shown on the color axis using Wilks' theorem.
Low values of $\beta$, corresponding to steep $dN/dS$, would imply a statistical excess of high-significance fluctuations below our detection threshold, distorting the TS distribution.
Higher values of $\beta$ would imply higher-peak-flux sources than observed.
The dashed line at $\beta = -2.5$ indicates the expectation for a Euclidean source distribution.
}
\label{fig:popstats}
\end{figure}

We formalize this by modeling the TS distribution (Fig.~\ref{fig:tshist}) as the sum of two parts: noise fluctuations and an injected population of simulated sources.
Noise fluctuations are based on realizations of the map noise at points uniformly distributed in the survey field (red line in Fig.~\ref{fig:tshist}).
The simulated sources are injected according to a power-law distribution of peak fluxes ($dN/dS \propto S^\beta$) and are likewise uniformly distributed throughout the survey region and summed with simulated noise at that point.
The peak times of the injected sources are scattered uniformly within the survey period at all points within 2 days of an SPTpol observation used in this analysis. 
For this test, each source is injected with a Gaussian profile (equation~\ref{eq:gaussprofile}) with the FWHM of the emission ($w$) set to one week, to approximately match the candidate source and the beginning of our roughly constant sensitivity region (Fig.~\ref{fig:gausssens}).
We run our normal search likelihood on each of these points and accumulate the TS value from the optimization of equation~\eqref{eq:ts}.
We then compare the resulting distributions of expected positive and negative fluctuations to the data (Fig.~\ref{fig:tshist}) in the region $\mathrm{TS} > 14.5$ using a Poisson likelihood.
This region corresponds to approximately $3.5\sigma$ and larger fluctuations, approximately where we would expect an astrophysical source population to be visible and in the high-TS asymptotic region that we have modelled and verified elsewhere in the analysis (Sec.~\ref{subsec:flaresearch}).
The resulting contours in the number of injected sources, normalized to the number with peak flux $\geq 20\, \mathrm{mJy}$, and $dN/dS$ index $\beta$ are shown in Fig.~\ref{fig:popstats}.

This fit prefers a non-zero total population largely because of the candidate object.
The results are dominated by the high-significance tail above the noise fluctuation background and so $dN/dS$ is poorly constrained given a constant number of above-threshold sources.
A profile likelihood calculation leaving the $dN/dS$ power-law index $\beta$ as a free parameter provides a source density above 20~mJy of $0.008^{+0.014}_{-.006}$ deg$^{-2}$ year$^{-1}$ and rejects zero at $2\sigma$.
Given the week-long sources injected, this corresponds to a snapshot density of  $1.5^{+2.7}_{-1.1} \times 10^{-4}$ deg$^{-2}$.
Removing the candidate source by hand gives a 90\% CL upper limit of $0.024$ deg$^{-2}$ year$^{-1}$ (snapshot rate $< 4.6 \times 10^{-4}$ deg$^{-2}$).

Theoretical expectations can provide similar numbers to the results obtained here, although large uncertainties on both the results of this analysis and the theoretical predictions, as well as the limited amount of theoretical work in this band, prevent any strong conclusions.
Using the model from \cite{ghirlanda14}, extended to 150~GHz (priv.\ comm.\ Ghirlanda), predicts $\sim 0.3$ orphan GRB afterglow detections in this survey and a $dN/dS$ index $\beta = -2.8$, compatible with the results here, although the emission period for these sources is expected to be much longer than the week emission of the candidate object.
Other predictions, such as in \cite{metzger15}, which consider different source populations (magnetars, on-axis GRBs, tidal disruption events), give expected source densities an order of magnitude lower.

As this is the first transient survey in the millimeter band, comparison of these results to previous surveys (e.g. \citealt{levinson02,galyam06,bower10,bower11,bell11,croft11}) is a complex and model-dependent task.
The highest frequency of these \citep{bower10} was conducted at 5~GHz, a factor of 30 below our primary observing band, and correspondingly was focused on somewhat different sources.
The most similar previous results in terms of science goals, in \cite{levinson02} and \cite{galyam06}, focus on GRB afterglows, though at lower frequencies (1.4~GHz).
These searches used two epochs of NVSS and FIRST data several years apart to identify the isotropic emission from a GRB afterglow after the ejecta become subrelativistic.
This occurs late in the history of the burst, with the peak time occurring six months or more after the burst and emission lasting for a period of a year \citep{levinson02,ghirlanda13}, a timescale to which the 1-year SPTpol survey described here has extremely limited or no sensitivity (Sec.~\ref{sec:sensitivity}).
Additional data from the in-progress 4-year 500 square degree SPTpol survey (Sec.~\ref{sec:discussion}) will allow a direct comparison of the results from \cite{levinson02} to millimeter-band data on similar timescales and corresponding limits on the allowed spectral index of sources like the possible detection in \cite{galyam06}.
The more indirect comparison---connecting constraints on late-time isotropic afterglows to the constraints we place on early partially beamed bursts---is highly theoretically uncertain for the reasons described in Sec.~\ref{sec:sensitivity} and an interesting topic for future modelling work.

While a direct comparison in terms of the year or more emission period implied by the \cite{levinson02} sub-relativistic afterglow model is impossible as a result of the length of our survey, the two-epoch strategy used sets only an upper bound on the length of their detected sources.
As such, we are at least free to compare results for week-scale sources and address whether the candidate source in \cite{levinson02} could be a similar object to that described in Sec.~\ref{sec:source}.

The similar effective sky coverage ($\sim 5000$~deg$^2$) for week-scale emission at fluxes $\gtrsim 10$~mJy suggests that any such sources detectable by both surveys at threshold must have a very flat broadband spectral index.
A reasonable synchrotron spectrum ($\nu^{-0.5}$) would make a week-long source detectable by \cite{levinson02} an order of magnitude below threshold here.
Conversely, even a very slowly rising spectrum such as expected for GRB afterglows near peak ($\nu^{0.3}$) would make an SPTpol-detectable source a factor of four below threshold for the FIRST/NVSS data; a steeper self-absorbed synchrotron spectrum ($\nu^2$) would make such sources completely invisible at 1.4 GHz.

This requirement for an extremely flat spectrum over two orders of magnitude in frequency makes it very unlikely that the candidate from \cite{galyam06} is related to the candidate in Sec.~\ref{sec:source}.
The population of bright ($\gtrsim 1$~mJy) GRB afterglows is dominated, both at 1~GHz and 150~GHz, by nearly on-axis self-absorbed bursts with rising spectra \citep{ghirlanda14}, which would make the majority of NVSS/FIRST-detectable afterglows visible to SPTpol given sufficiently long observation times.
Future SPTpol data over longer time periods will thus allow a much more direct comparison to the results from \cite{levinson02}.

\section{Discussion and Future Work}
\label{sec:discussion}

Observations with SPTpol have provided the first untriggered view of the transient millimeter sky, with sensitivity approaching that required to test current models of off-axis GRBs and other sources.
One candidate object was observed, but it remains unclear whether the observed emission is a statistical fluctuation.
Its properties are intriguing and qualitatively consistent with some expectations for a GRB afterglow, although there is some internal tension between the polarization data and the spectrum and the statistical significance of the detection is too low to completely rule out a fluctuation.

The forthcoming SPT-3G receiver, scheduled for deployment over the 2016/2017 austral summer \citep{benson14}, will greatly improve the capabilities of this survey and provide much-enhanced sensitivity to any objects with the same properties (peak flux, duration, spectrum) as the candidate from this work.
SPT-3G is planned to survey an area of 2500 deg$^2$---25 times the area covered here---in three frequency bands (90, 150, and 220~GHz) to approximately the same depth as in this article, ultimately probing source densities in the 10--20~mJy regime to below $10^{-4}$ deg$^{-2}$ year$^{-1}$.
At this level, multiple detections are expected annually from the off-axis bursts modeled in \cite{metzger15} and \cite{ghirlanda14}.
Even a non-observation with SPT-3G will thus place constraints on the shock dynamics and energy budget of the unknown GRB progenitors.
Most important, this sensitivity is well below the source density implied by the candidate source here.
If it was a statistical fluctuation, SPT-3G will be able to rule out a transient source population at the best-fit level in Fig.~\ref{fig:popstats}.
Conversely, if it was indeed a real source, SPT-3G would see dozens of sources annually at our best-fit source density, independent of $\beta$, and begin to characterize the population from which it arose.

In addition, a 500 deg$^2$ survey using SPTpol is currently in progress with a planned conclusion at the end of 2016.
For week-scale sources, this survey will cover an effective sky area 20 times larger than covered here, albeit with map noise approximately two times greater.
This prevents the direct exploration of the population of 10~mJy sources possible with SPT-3G, but will provide complementary information to what is available in this work.
For a source population with $dN/dS$ index $\beta > -3$ such as predicted for all source classes in \cite{metzger15}, including orphan GRB afterglows, the number of detections will increase by trading depth for sky area.

Extending the observing period from one year to the four years of the 500 square degree survey will also allow better discrimination between dim steady sources and long transients.
Orphan GRB afterglows in particular, as well as population-3 GRBs, may have durations of months.
With only one year of data, such objects are largely indistinguishable from steady sources, reducing the sensitivity of this analysis despite the low noise levels afforded by long integration times.
As a result, the 500 square degree survey is expected to have equivalent flux sensitivity on 100-day scales to the results here in addition to the substantially increased sky area and monitoring period.

\acknowledgements{
Thanks to S.~Croft for helpful comments and to G.~Ghirlanda for providing 150~GHz orphan afterglow predictions.
The South Pole Telescope program is supported by the National Science Foundation through grant PLR-1248097. Partial support is also provided by the NSF Physics Frontier Center grant PHY-0114422 to the Kavli Institute of Cosmological Physics at the University of Chicago, the Kavli Foundation, and the Gordon and Betty Moore Foundation through Grant GBMF\#947 to the University of Chicago for the construction of SPTpol.
The McGill authors acknowledge funding from the Natural Sciences and Engineering Research Council of Canada and Canadian Institute for Advanced Research.
JWH is supported by the National Science Foundation under Award No. AST-1402161.
BB is supported by the Fermi Research Alliance, LLC under Contract No. De-AC02-07CH11359 with the U.S. Department of Energy.
TdH is supported by the Miller Institute for Basic Research in Science.
The CU Boulder group acknowledges support from NSF AST-0956135.
This work is also supported by the U.S. Department of Energy.
Work at Argonne National Lab is supported by UChicago Argonne, LLC, Operator of Argonne National Laboratory (Argonne).
Argonne, a U.S. Department of Energy Office of Science Laboratory, is operated under Contract No. DE-AC02-06CH11357.
We also acknowledge support from the Argonne Center for Nanoscale Materials.
The data analysis pipeline uses the scientific python stack \citep{hunter07, jones01, vanDerWalt11} and the HDF5 file format \citep{hdf5}.
Computing for this work was conducted using resources provided by the Open Science Grid \citep{pordes07}, which is supported by the National Science Foundation and the U.S. Department of Energy's Office of Science.
}

\newcommand{\pasa}{Proc. of the Astronomical Society of Australia}

\end{document}